%% file: main.tex
\documentclass[10pt,onecolumn ]{article}
\usepackage{hyperref}
\usepackage[english]{babel}
\usepackage{appendix}
\usepackage{graphicx}
\usepackage{wrapfig}
\usepackage{amsmath}
\usepackage{authblk}
\usepackage{subfiles}
\graphicspath{ {figures/} }

\begin{document}
\title{Quantum emitters coupled to circular nanoantennas for high brightness quantum light sources}
\author[1]{Hamza A. Abudayyeh}
\author[1,2]{Ronen Rapaport}
\affil[1]{ Racah Institute for Physics, and}
\affil[2]{ Applied Physics Department, The Hebrew University of Jerusalem, Jerusalem 9190401, Israel}
\date{}

\maketitle

\begin{abstract}
Engineering the  directionality and emission rate of quantum light sources is essential in the development of modern quantum applications. 
In this work we use numerical calculations to optimize the brightness of a broadband quantum emitter positioned in a hybrid metal-dielectric circular periodic nanoantenna.  
The optimized structure features a photon collection efficiency of $74\%$ ($82\%$) and a photon flux enhancement of over 10 (6) into a numerical aperature of 0.22 (0.50) respectively, corresponding to a direct coupling into two types of multimode fibers. 
In order to enhance the emission rate, we present a new circular nanoantenna design where a quantum emitter is attached to a silver nanocone at the center of the antenna. After optimization, we find a collection efficiency of $61\%$ ($78\%$)  into a numerical aperature of 0.22 (0.50), giving a brightness enhancement of 1000 (600) for an unpolarized emitter.
The enhancements in both structures are broadband due to the low quality factor of the device and are therefore ideal for room-temperature sources.
This type of a scalable design can be utilized towards on-chip, high brightness quantum light sources operating at room temperature.
\end{abstract}

\section{Introduction}

Single quantum emitters are at the heart of various quantum applications  such as quantum computation, encryption, simulations, communications, metrology among many others.\cite{OBrien2009PhotonicTechnologies,Dowling2003QuantumRevolution} 
As candidates for such applications, many quantum emitters have been studied \cite{Lounis2005} including trapped atoms \cite{Kimble1977PhotonFluorescence}, single molecules \cite{Treussart2001PhotonFilm},  impurity centers (e.g silicon vacancies (SiV) in diamond \cite{Neu2012PhotophysicsEmission}), nanocrystal colloidal quantum dots (NQDs) \cite{Lounis2000PhotonFluorescence}, and self assembled quantum dots \cite{Ding2016}. 

Two limitations face free-standing quantum emitters however. The first lies in their isotropic angular emission pattern which limits the ability to efficiently collect their emitted photons. 
The second is that typical solid state quantum emitters can have low emission rates which will limit the speed of future technologies. 
In the past decade there have been considerable efforts to solve these deficiencies by modifying the photonic environment near the quantum emitter. \cite{Dey2016PlasmonicDots,Pelton2015ModifiedStructures}
To achieve this, emitters were embedded in, or near to various nanostructures.  For simplicity, in this work we will refer to any nanostructure that modifies the emission rate or directionality of a quantum emitter as an antenna.

One approach for improving the directionality and emission rate of quantum emitters is the use of metallic antennas.  These include metal nanoparticles (MNPs) \cite{Dey2016PlasmonicDots}, plasmonic patch antennas \cite{Esteban2010OpticalResonances,Belacel2013ControllingAntennas,Bigourdan2014DesignEmission}, metallic nanoslit arrays \cite{Livneh2011}, Yagi-Uda nanoantennas \cite{Curto2010UnidirectionalNanoantenna,Dregely20113DArray} and circular bullseye plasmonic nanoantennas \cite{Li2013ActiveBeams,Harats2014}. 
The advantage of using  such structures is that plasmonic modes have low mode volumes accompanied with low quality factors enabling spontanous emission 
lifetime shortening and emission redirection over broad spectral ranges which is beneficial for room-temperature broadband sources such as colloidal quantum dots and color centers in diamond \cite{Giannini2011PlasmonicNanoemitters}. 
On the other hand, for efficient coupling the emitter has to be placed in the vicinity of the plasmonic structure which will increase nonradiative recombination rates and if placed too close will cause quenching of the emission all together \cite{Dey2016PlasmonicDots} .

Another approach is to use pure dielectric antennas such as microcavities \cite{Ding2016} and photonic crystals \cite{Englund2009QuantumOptics,Laucht2012BroadbandWaveguides,MangaRao2007SingleWaveguide} that feature high directionality, high radiative enhancement factors, and low-loss \cite{Ates2009,Davanco2011AEmission}. 
Despite these advantages however, dielectric antennas usually come with a limiting narrow frequency bandwidth that is usually unsuitable for room temperature quantum emitters \cite{Krasnok2016DemonstrationStructures}, and are much more complex to fabricate.

Therefore a hybrid metal-dielectric nanoantenna that combines the advantages of metallic and dielectric antennas but with much less of their drawbacks was developed. In such a design, the emitter can be placed at a large distance from the metal and still produce high directionality in a broad spectral range \cite{Livneh2015EfficientNanoantenna,Livneh2016}. 
Our recent experiments demonstrated that the single photons emitted from a single nanocrystal quantum dot (NQD) positioned in such a hybrid circular ('bullseye') antenna can be collected with an efficiency of around $37 \%$ into a moderate numerical aperture (NA) of 0.65. Furthermore, the ability of positioning a single NQD at the hotspot of the antenna was developed which enables fabrication of highly directional room-temperature single photon sources \cite{Livneh2016,Harats2017DesignEmission}.

While this points into a promising direction the system should be further optimized to enhance both the collection efficiency and decay rate of a quantum emitter.
This optimization is crucial for applications such as quantum key distribution, single shot readouts of nitrogen vacancy (NV) centers, sensing applications, and random number generation \cite{Rarity1994QuantumSharing,Lounis2005,Wolf2015Purcell-enhancedDiamond,Aharonovich2016Solid-stateEmitters} some of which will be discussed further below.
It is therefore pivotal to develop an antenna that combines both high collection efficiencies and high decay rate enhancement. 
Another drawback of the previous design was the plasmonic emission due to the metal corrugation near the quantum emitter which reduced the single photon purity of the source. \cite{Livneh2016}
Consequently it would be desirable, for single photon sources at least, that the central area lying within the focus of the exciting laser be flat. 

In this work we introduce a new hybrid circular antenna where the emitter is positioned in proximity to a central metal cone at the center of the bullseye antenna, and use a finite-difference time-domain (FDTD) method \cite{Taflove2005ComputationalMethod,LumericalInc.} to optimize the brightness of the emitter into low NA's, with the goal to have a high brightness broadband source of single photons directly coupled to a fiber, without the need of any additional collimating optics. This is achieved by simultaneously optimizing the emission directionality and the enhancement of the emission rate of a dipole emitter. The particle swarm algorithm \cite{Robinson2004ParticleElectromagnetics} was used for the optimization process.  

In section 2 we define relevant quantities to be optimized, such as the collection efficiency and the enhancement factors of the emitters brightness. In section 3 and 4 we will introduce two optimized designs for a bullseye antenna without and with a central enhancing element respectively.

\section{General definitions: coupling between a dipole emitter and an antenna}
The decay rate of a dipole emitter $\Gamma$ in an inhomogeneous environment will depend on the orientation of the dipole moment with respect to the polarization of the electromagnetic modes. 
Thus in general the response of an antenna will vary for different dipole orientations and therefore we will label all subsequent quantities with an index $i= x,y,z$ that represents the dependence on the dipole orientation with respect to some coordinate system.
For an emitter in free space the local density of states is isotropic and the decay rate reduces to $\Gamma_{0i}$.
This intrinsic decay rate in general includes both radiative ($\Gamma_{0i}^r$) and non-radiative ($\Gamma_{0i}^{nr}$) decay rates.
Intrinsic non-radiative loss mechanisms vary for different emitters and include for example phonon-scattering \cite{Jagtap2015ExcitonphononDots}, Auger recombination of multiexcitons in colloidal quantum dots, \cite{Robel2009UniversalNanocrystals}, and losses to lattice defects in color centers \cite{Neu2012PhotophysicsEmission}.

In a modified environment the decay rate ($\Gamma_i$) changes and the Purcell factor is defined as \cite{Purcell1946SpontaneousFrequencies}:
\begin{equation}
\label{eq: Purcell}
F_i=\frac{\Gamma_i}{\Gamma_{0i}}.
\end{equation}
where $F_i =F(\mu_i)$ and $\Gamma_i=\Gamma(\mu_i)$ and so on represents the dependence of these quantities on the dipole orientation as discussed above.
This factor determines the reduction ($F_i>1$) or increase ($F_i<1$ ) of the intrinsic lifetime of the emitter due to the modified electromagnetic environment and in general does not equal the enhancement of the emitter's emission into the far-field since part of the emission can be channeled into non-radiative losses in the antenna (such as Joule losses in metals). 
Thus we define a radiative enhancement factor $F^r_i$ as the net increase of total photon flux while the non-radiative enhancement factor $F^{nr}_i$ depicts the leakage of emission into lossy modes of the antenna (therefore modifying the non-radiative decay rate). 
Note that the antenna in most cases doesn't affect the intrinsic non-radiative decay rate and thus we can write the total decay rate as: 
\begin{equation}
\label{eq: Gammadef}
\Gamma_i=F_i\Gamma_{0i} = \underbrace{F^r_i \Gamma_{0i}^r}_{\Gamma_i^r} +\underbrace{F^{nr}_i \Gamma_{0i}^{r}+\Gamma_{0i}^{nr}}_{\Gamma_i^{nr}}
\end{equation}

In free space the intrinsic quantum yield $QY_{0i} =\Gamma_{0i}^r/\Gamma_{0i}$ defines the probability that the photon is emitted into the far-field. 
The presence of an antenna in general will modify the radiative and non-radiative decay rates and therefore the overall quantum yield of an antenna-emitter system $QY_i=\Gamma_i^r/\Gamma_i$  is altered and can be written as: 
\begin{equation}
\label{eq:QY}
QY_i=\frac{F^r_i}{F_i}QY_{0i}
\end{equation}
The effect of the antenna on the quantum yield depends on the intrinsic features of the dipole ($\Gamma^r_{0i},\Gamma^{nr}_{0i}$) and antenna ($F^r_i,F^{nr}_i$) in a coupled fashion since $F_i$ is defined as in equation \ref{eq: Gammadef}.
In this work we will be considering a dipole emitter with unity intrinsic quantum yield and therefore $F_i$ reduces to $F_i^r+F_i^{nr}$ and we can define an antenna quantum efficiency $AE=F^r_i/F_i$ which is independent of the properties of the dipole emitter.

Another interesting property of the emitter-antenna system is the far-field brightness. An emitter with an intrinsic radiative decay rate ($\Gamma^r_{0i}$) will have a total far field brightness denoted by $\Phi^{tot}_{0i}$ ($tot$ refers to integration over all angles), that is dependent on the excitation power.
We define the brightness of the emitting system as the flux that is emitted into a certain collection cone of given $NA$, given by:

\begin{figure}[t!]
\centering{\includegraphics[width=0.5 \textwidth]{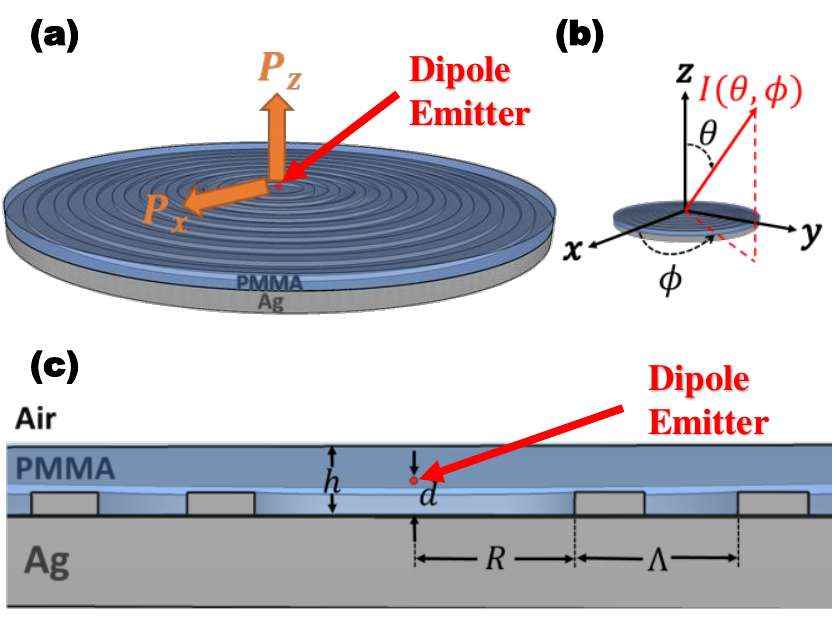}}
\caption{(a) Schematic representation of the hybrid metallic dielectric bullseye nanoantenna with empty central area displaying the two independent dipole orientations. (b) The azimuthal ($\phi$) and polar ($\theta$) angles of the far field intensity distribution. (c) Cross section of the nanoantenna defining the key geometric parameters: the central cavity radius ($R$), groove period ($\Lambda$),    emitter height ($d$), and polymer thickness ($h$) }
\label{fig:1}
\end{figure}
\begin{equation}
\label{eq: Flux}
\Phi_i(\theta_{NA})=\Phi^{tot}_{0i} F^r_i \eta_i(\theta_{NA})
\end{equation}
where $\eta_i(\theta_{NA})$ is the collection efficiency which quantifies the directionality of the emitting system and can be written in terms of the flux per unit solid angle or intensity $I_i(\theta,\phi)$ as:
\begin{equation}
\label{eq: CollEff}
\eta_i(\theta_{NA})=\frac{\int_0^{2\pi}d\phi\int_0^{\theta_{NA}} d\theta\ I_i(\theta,\phi)}{\int_0^{2\pi}d\phi\int_0^{\pi} d\theta\ I_i(\theta,\phi)}
\end{equation}
It is also useful to define the brightness enhancement, $\xi_i(\theta_{NA})$,  as the ratio of the brightness of a dipole emitter into a certain NA to the brightness of a free dipole emitter into the same NA, i.e.:
\begin{equation}
\label{eq: BrightnessEnhancement}
\xi_i(\theta_{NA})= \frac{\Phi_i(\theta_{NA})}{\Phi_{0i}(\theta_{NA})} =  F^r_i \frac{\eta_i(\theta_{NA})}{\eta_{0i}(\theta_{NA})}
\end{equation}
where $\eta_{0i}(\theta_{NA})$ refers to the collection efficiency of a free dipole. 

The coupling of an emitter to an antenna has a few important consequences as shown in equations \ref{eq:QY} and \ref{eq: Flux}. 
Firstly, the quantum efficiency of the emitter or more generally the quantum efficiency of certain emission processes of the quantum emitter will be modified. For example it has been shown that  coupling colloidal quantum dots with metal nanoparticles will increase the biexciton quantum efficiency \cite{Matsuzaki2017StrongAntenna}. 
This could be detrimental or beneficial depending on the application (e.g., single photon sources vs. bi-photon sources).
The other two consequences are the modification of the emission pattern and decay rate which are the main focus of this paper.

\section{Circular periodic ('Bullseye') nanoantenna with a central cavity}
The structure studied in this section is based on our previous work in Refs \cite{Livneh2016,Harats2017DesignEmission}, as is shown schematically in figure \ref{fig:1}c and d. 
It consists of a bullseye circular grating with 15 gratings,  period $\Lambda$ and a central cavity of radius $R$ embedded in a dielectric layer of thickness $h$ (realized by a polymer in \cite{Livneh2016,Harats2017DesignEmission}). The emitter is considered as an oscillating dipole at the center of the circular grating and at height $d$ from the metallic slab. 

The large central cavity in the current antenna is introduced since we previously found that the presence of metallic corrugation near the emitter leads to efficient excitation of surface plasmons on the metal-dielectric interface which then results in a noticeable broadband random  emission of photons, resulting not from the quantum emitter itself but rather from these surface plasmons. This excess random emission degraded the performance of a single photon device \cite{Livneh2016}. 

\begin{figure}[t!]
\centering{\includegraphics[width=0.5 \textwidth]{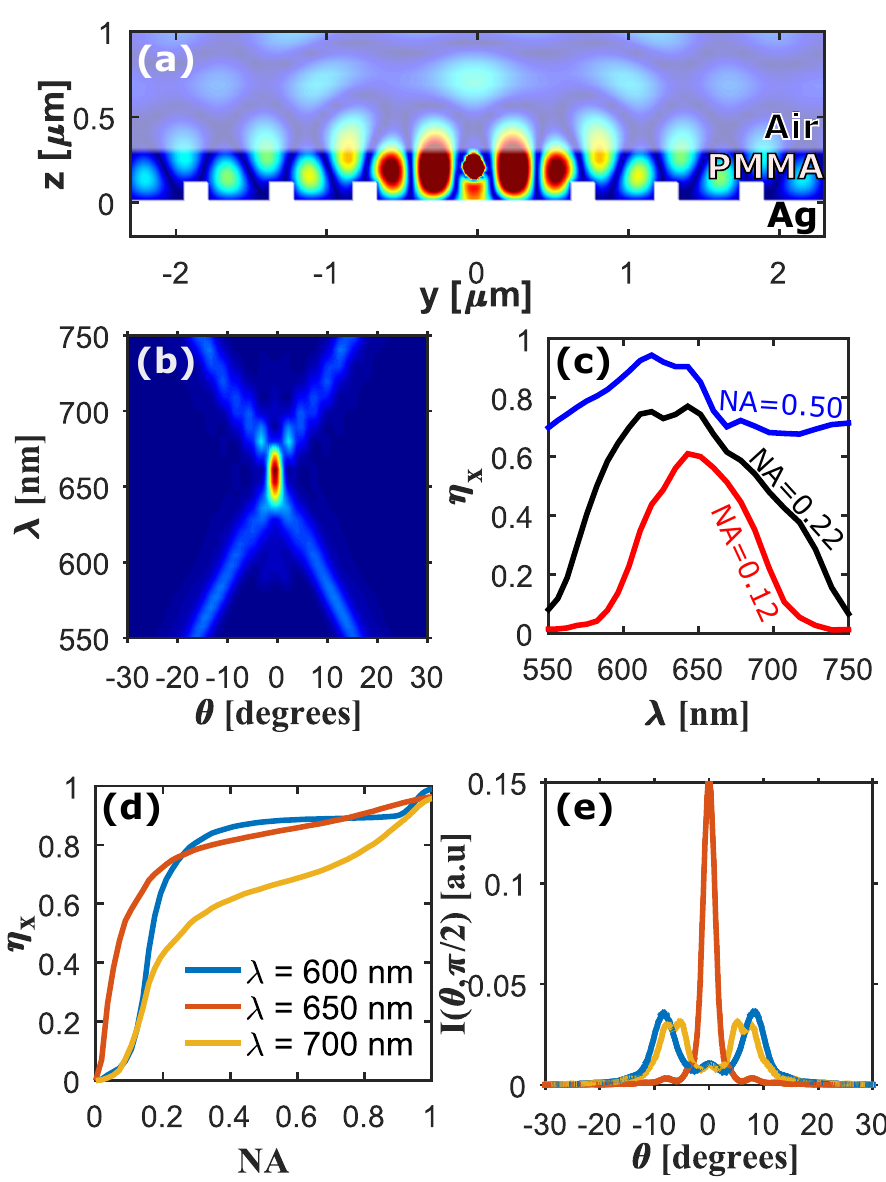}}
\caption{ 3D Lumerical\textsuperscript{\textregistered} simulation of the Bullseye Nanoantenna. (a) Electric field distribution due to a dipole emitter oriented along the x axis taken in the yz plane. (b) Angular emission spectrum $I(\theta,\lambda)$  for $\phi=\pi/2$  (yz-plane). (c) Collection efficiency into different NA's  vs wavelength displaying the broadband response of the system. Collection efficiency (c) and angular intensity distribution for $\phi=\pi/2$ (d) for an emitter on resonance $\lambda = 650 nm$ (red), and detuned from the antenna resonance $\lambda = 600 nm$ (blue) and $\lambda = 700 nm$ (yellow).  }
\label{fig:2}
\end{figure}
The operation mechanism of this system is as follows: When a quantum emitter (having a spectral band around the central emission wavelength $\lambda_0$) at the center of the bullseye structure is excited it emits photons preferentially into the dielectric layer due to the higher index of refraction. 
The dielectric layer is designed to have modes that support this band of wavelengths (see supplementary information A) with corresponding propagation constants $\beta(\lambda)$. 
As the photons propagate in the radial direction they encounter a diffraction grating with period $\Lambda$ that is designed to match the propagation constant ($\beta \approx 2\pi/\Lambda$) which diffracts the light.
The diffracted photons that hit the air-dielectric surface with angles less than the critical angle couple out while all other photons continue to the next grating. 
Eventually nearly all the light is diffracted into free space modes and the interference between the various diffracted beams leads to a low divergence beam in the far field. 

In addition to the geometric parameters of the antenna, the dipole orientation has a profound role in the performance of the device.
Due to the azimuthal symmetry of the antenna we only need to consider two independent dipole orientations, the in-plane dipole ($P_x$)  which will be represented by a dipole  along the x axis (see figure \ref{fig:1}b)  and the out of plane dipole ($P_z$ ).
Since the $P_z$ dipole has a polarization that is predominantly in the z direction, it will mostly couple to the z component of the $TM_0$ electric field for the metal-dielectric-air (MDA) slab waveguide (see supplementary information A). 
The electric field of this component is maximum near the metal surface and thus coupling into this mode requires either thicker dielectric layers which introduces higher order modes, or placing the dipole closer to the metal where diffraction is limited to the first few gratings thus lowering the redirection effect caused by constructive interference. 
On the other hand the $P_x$ dipole couples into the $TE_0$ and the x-component of the $TM_0$ mode which both have field maxima near the center of the dielectric layer. This leads to better constructive interference into lower angles due to the diffraction caused by multiple circular grooves. 
It is clear from this discussion and from our simulations that the emission from the $P_x$ dipole will have a better coupling with the antenna and thus better directionality. 
Moreover due to azimuthal symmetry an unpolarized source will have a probability of $2/3$ to emit from an in-plane dipole (see supplementary information B). 
For these reasons we have chosen to optimize the structure considered in this section for the $P_x$ dipole.

The performance of the device will be contingent on the geometrical parameters of the nanoantenna (see figure \ref{fig:1}d) and the dipole orientation. 
Hence an optimization for these parameters for the $P_x$ dipole was conducted using a particle swarm algorithm \cite{Robinson2004ParticleElectromagnetics} in a  3D FDTD Lumerical\textsuperscript{\textregistered} simulation \cite{LumericalInc.}. 
The figure of merit for this optimization procedure was the system's brightness, i.e., its total flux, into an $NA = 0.22$   ($\Phi_x(0.22)$) in equation \ref{eq: Flux}, corresponding to a typical NA of multimode fiber.
The total flux was chosen rather than the collection efficiency to incorporate the radiative enhancement factor.
The parameters are all normalized to the central wavelength of the source ($\lambda_0$) since the optimization shows that the structure scales well with wavelength in the spectral range between 600 and 800 nm (see supplementary information C). The dipole emitter was chosen to have a central wavelength of $\lambda_0$ = 650 nm.
The optical parameters of the dielectric layer was chosen to match that of PMMA (Polymethyl methacrylate) which is optically transparent in the desired spectral range and is used for standard fabrication procedures \cite{Harats2017DesignEmission}. 

The optimized structure described in this section has a normalized central radius $R/\lambda_0 = 0.97$, normalized grating period $\Lambda/\lambda_0 = 0.86$, normalized dielectric layer thickness $h/\lambda_0 = 0.41$, and normalized dipole height $d/\lambda_0 = 0.29$. The normalized ring width and height were set to $0.25$ and $0.15$ respectively. 

From the simulations we can visualize both the near-field and far-field distributions. Figure \ref{fig:2}a displays the simulated electric field inside the optimized structure in the plane normal to the dipole orientation (yz-plane) .
In this plane the emission couples into the $TE_0$ mode of the waveguide and we can observe that the standing waveguide modes have effective wavelengths that match the periodicity of the gratings.
In the air layer above, the fields propagating out of the structure are visible. 
These fields have phases resulting in constructive interference in the far field  into a narrow cone about $\theta =0$ .
The far field spectral response of the antenna for wavelengths around the resonance wavelength ($\lambda = 650$ nm) is displayed in figure \ref{fig:2}b. 
This far-field  spectral intensity distribution ($I(\theta,\lambda)$) is plotted for the yz-plane ($\phi=0$). Angular cross sections for the resonance wavelength ($\lambda=650$ nm) and two detuned wavelengths ($\lambda= 600$ and $700$ nm) are plotted in figure \ref{fig:2}e.  
From these figures we can clearly see the dispersion resulting from the circular grating as the first orders ($\pm 1$) combine around $\lambda = 650$ nm to form a narrow directional beam.
To better quantify the directionality of the emission the collection efficiency (defined in equation \ref{eq: CollEff}) for the same three wavelengths is shown in figure \ref{fig:2}d. 
In figure \ref{fig:2}c the collection efficiency as a function of wavelength is displayed for three collection NA's (0.12,0.22,0.50) corresponding to the NA of a single mode fiber and two commerically available multimode fibers respectively.
From this figure and  the overlap of the modes in figure \ref{fig:2}b it is evident that the antenna's response is broadband ($> 20$ nm)  which is beneficial for coupling to a room-temperature emitter. 
It is clear that detuning the emitter's wavelength from the designed wavelength will result in a lower collection efficiency at low NA corresponding to the formation of sidebands as seen in figure \ref{fig:2}e.
\begin{figure}[t!]
\centering{\includegraphics[width=0.5 \textwidth]{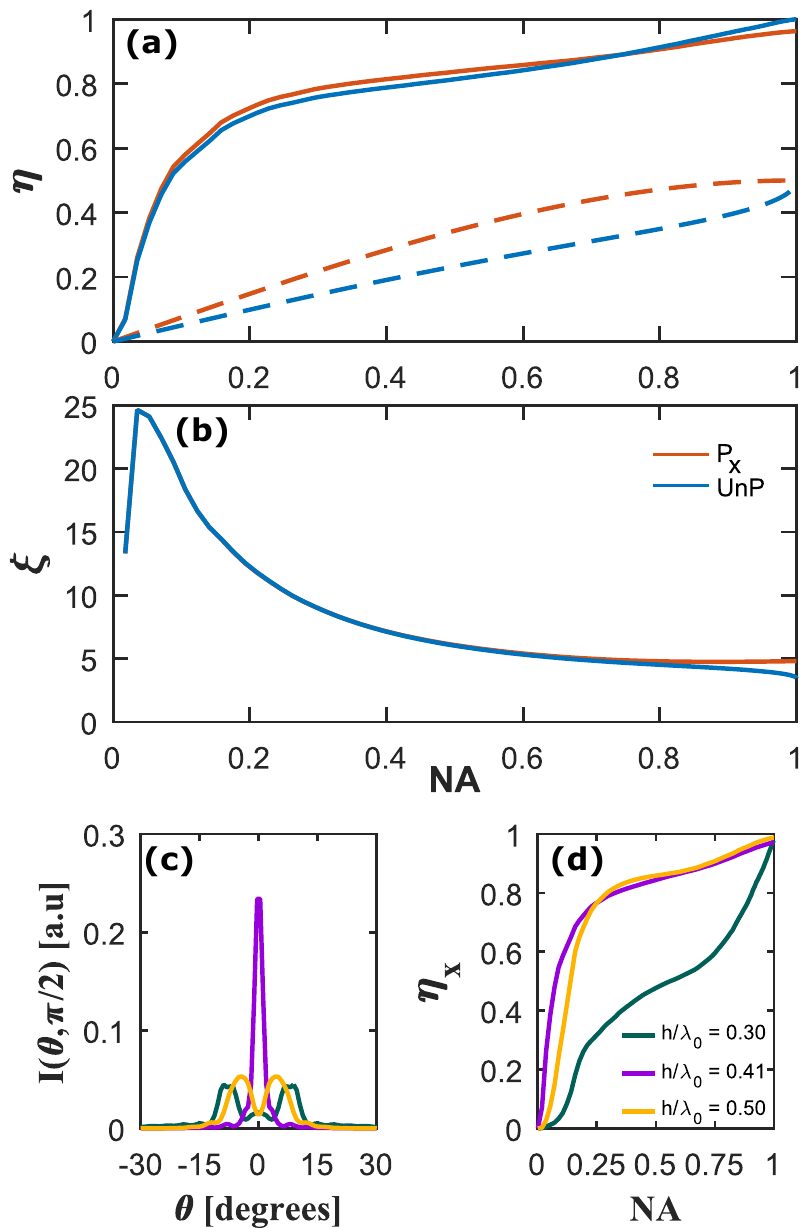}}
\caption{Collection efficiency (a) and brightness enhancement (b) for a dipole oriented along the x axis ($P_x$) and for a dipole emitting in a random orientation ($UnP$).  The solid lines represent dipoles coupled to the antenna and the dashed lines represent dipoles in free space. Angular intensity distribution (c) and collection efficiency (d) for the optimized polymer thickness ($h/\lambda_0 =0.41$) and for a thinner ($h/\lambda_0 = 0.3$)  and thicker ($h/\lambda_0 =0.5$) waveguide thickness. }
\label{fig:3}
\end{figure}

Next we turn our attention to evaluating the performance of the device in general and in particular to the dependence on dipole orientation. 
Due to the low-loss of the device we will be interested in the collection efficiency (eqaution \ref{eq: CollEff}) and brightness enhancement (equation \ref{eq: BrightnessEnhancement}).
In this convention the radiative enhancement factor $F^r_i $ is just $\xi_i(\pi)/2$. 
The collection efficiency and the brightness enhancement for a dipole oriented along the x-axis ($P_x$) and for an unpolarized dipole ($UnP$) are displayed in figure \ref{fig:3}a and b respectively. 
These are compared to the corresponding values for free dipoles (dashed curves). 
The brightness for the unpolarized emitter is calculated from the two polarized ones, $P_x$ and $P_z$ (not shown here) as discussed in the supplementary information B. 
Using this brightness one can directly calculate the collection efficiency. 

One very promising result is that around $74\%$ of the emission of both the $P_x$ and unpolarized emitters is collected already into a low, multi-mode fiber compatible numerical aperture of $NA=0.22$, and an even higher collection efficiency of $82\%$ is found for $NA=0.5$, corresponding to a larger $NA$ commercially available multi-mode fiber.
This high coupling efficiencies into such low NA's should enable direct coupling of the emission into a commercial multimode fiber. 
Moreover, as a result the confinement introduced by the antenna there is a net radiative enhancement of 2.4 and 1.7 for the $P_x$ and unpolarized emitters respectively, which together with the improved collection, results 
in a brightness enhancement (figure \ref{fig:3}b) of $11$ ($6$) into the corresponding NA's compared to that of a free standing dipole. This means a large increase in the usable photon rate from the source.
It is visible that the brightness enhancement is especially high for low NA's which is understandable due to the poor collection efficiency from the free-standing dipoles at low collection angles. 
It is worth noting that the brightness enhancement of the $P_x$ and unpolarized dipoles overlap completely for nearly all the collection angles.

Another important point is the near identical collection efficiency curve of the $P_x$ and unpolarized dipoles. 
This is due to the enhancement of the emission of $P_x$ dipole and suppression of the $P_z$ dipole so that the unpolarized dipole emission will result primarily from an in-plane dipole especially for low collection angles. 
This has the profound effect of transforming an emitter that originally emitted from all dipole orientations to an emitter that primarily emits from a dipole in a plane perpendicular to the nanoantenna. 
There is however, a probability of $1/3$ to emit from a $P_z$ dipole during each excitation cycle and the longer lifetime of this transition will result in a less time-deterministic source.
Therefore it is advantageous to use a source that is originally polarized and whose dipole can be oriented in the plane of the nanoantenna. 
Such a source will enable the full use of all the benefits of this design while maintaining the deterministic nature of the source. 
One interesting candidate is the colloidal quantum rod  reported by Sitt et al. \cite{Sitt2011HighlyPolarization}
. In these quantum rods the symmetry is broken by using elongated shells (dot in rod) or elongated cores and shells (rod in rod) and the preferred transition has a dipole moment along the long axis of the quantum rod yielding degrees of polarization reaching $83\%$. \cite{Hadar2013,Sitt2012}. When spin-coated, the rods align in plane making them ideal for implementation in this new antenna design. 
Recently there has also been some significant progress in controlling the orientation of the emission dipole in NV centers in diamond during chemical vapor deposition growth \cite{Schroder2016QuantumInvited}, which can also make them suitable for the proposed scheme.

Finally, we address the effect of varying the thickness of the waveguide layer.  The angular intensity distribution and the collection efficiency for three thicknesses namely $h/\lambda_0 = 0.3,0.41,$ and $0.5$ are shown in figure \ref{fig:3}c and d respectively. Slightly changing the waveguide thickness will change the propagation constant of the fundemental $TE_0$ and $TM_0$ modes. 
Increasing the thickness further will result in the contribution of higher order modes (see Supplementary Information A)
The former is the cause for the ranges of thicknesses in figure \ref{fig:3} c and d. 
The mismatch between the propagation constant and the grating Bragg wavevector leads to the formation of sidebands.  
The effect of the other geometrical parameters are discussed in the supplementary information section C. 

One particularly interesting implementation (and perhaps test) of such designs is for single shot readout of spin-states in nitrogen vacancy centers in diamond. It was theoretically established that moderate Purcell factors might lead to a significantly enhanced SNR. \cite{Wolf2015Purcell-enhancedDiamond}. 
As shown above such moderate Purcell factors can be realized. In fact, it is expected  that for diamond waveguides the Purcell factor will be higher than reported above due to the higher index of refraction which leads to tighter confinement. 
This is confirmed by initial calculations (not shown here) where radiative enhancement factors as high as 8 where achieved. 
Furthermore, the SNR of the readout of the spin-state also benefits from efficient collection of the emitted photons, which is an added advantage of this design 


It should be noted that the fabrication and analysis of bullseye structures around single NV centers has recently been reported with purely dielectric bullseye structure. \cite{Li2015EfficientGrating} The predicted collection efficiency however was only $13\%$ into an $NA=0.7$.
Other advances in the fabrication of photonic and plasmonic structures in diamonf have been recently reported such as plasmonic apertures  \cite{Choy2011EnhancedAperture} and periodic nanocrystal arrays \cite{Aharonovich2013Bottom-upNano-structures}. 
Silicon vacancies also provide for an interesting application particularly due to their near lifetime limited linewidths \cite{Sipahigil2014IndistinguishableDiamond,Li2016NonblinkingNanodiamonds}. Therefore implementing this design will open up an opportunity towards a highly efficient source of indistinguishable single photons necessary for quantum logic gates \cite{Aharonovich2016Solid-stateEmitters} 

\section{Composite Bullseye-nanocone nanoantenna}

As discussed before, in addition to high collection efficiency, it is desirable to enhance the decay rate of a quantum emitter. The design of the previous section had a rather small increase of the intrinsic radiative rate of the emitter, due to the poor confinement of the optical mode around the dipole, and the low LDOS of the essentially dielectric waveguide mode. 
To overcome this intrinsic limitation, we have considered introducing a plasmonic structure at the center of the bullseye nanoantenna. An interesting candidate is the silver or gold nanocone.
These nanocones have a longitudinal mode along the cone's axis and a transverse mode along the base. \cite{Meixner2015}
The longitudinal mode in particular offers high field enhancement due the lightning rod effect near the rod's tip.  \cite{Fleischer2008Three-dimensionalMicroscope}
\begin{figure}[t!]
\centering{\includegraphics[width=0.5 \textwidth]{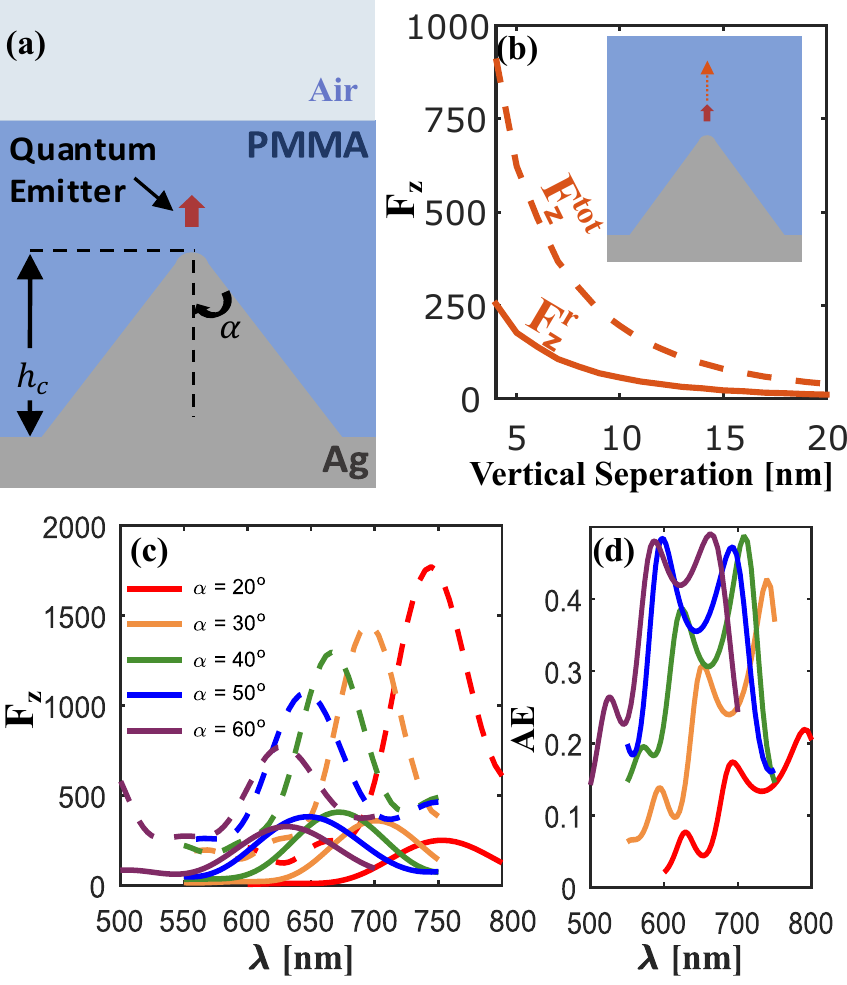}}
\caption{(a) Schematic for the a silver nanocone with rounded tip embedded in a dielectric layer. (b) Radiative (solid) and total (dashed)   enhancement factor as a function of vertical seperation between the dipole and nanocone with $\alpha = 50^o$ . (c) Radiative (solid) and total (dashed) decay rate enhancement  for nanocones of different apex angles. (d) Antenna quantum efficiency for nanocones of different apex angles }
\label{fig:4}
\end{figure}
Recent theoretical studies on silver nanocones showed that the plasmonic resonance can be tuned across the optical spectrum  by changing the apex angle of the cone. \cite{DAgostino2013DipoleNanocones,Goncharenko2006ElectricMicroscopy,Goncharenko2007ElectricNano-tip}
Furthermore the study demonstrated that the largest  quantum yields occur for larger nanocones of heights around $160$ nm and angles of $\pi/6$\cite{DAgostino2013DipoleNanocones}. 
A similar theoretical study on gold nanocones also displayed the shift of the plasmonic resonance with tip angle at fixed cone heights. \cite{Mohammadi2010FluorescenceAntenna}
A detailed study of the electric field enhancement of these nanocones and their dependence on the cone parameters can be found in, e.g., \cite{Martin2001StrengthMicroscopy,GoncharenkoElectricDependence,Goncharenko2006ElectricMicroscopy,Goncharenko2007ElectricNano-tip}

The difficulty of using nanocones for emission rate enhancement presides in the fact that the emitter must be placed with sub 10 nm precision near the tip of the nanocone. 
Furthermore, the vertical separation between the emitter and the tip needs to be of the order of a few nanometers due to the high localization of the field. 
Recently both problems were addressed by chemically binding NQDs preferentially to the tip of gold nanocones .\cite{Fulmes2015Self-alignedNanostructures}
The molecular binding and the shell of the NQD provided a vertical separation of around 4 nm. \cite{Fulmes2015Self-alignedNanostructures}.

Since we are interested in exciting the longitudinal mode of the nanocone a dipole oriented along the z-axis is considered in this section. 
First we consider a nanocone of height $h_c$ and apex angle $\alpha$ on a metal slab embedded in a dielectric (PMMA) layer without the presence of the bullseye rings (see figure \ref{fig:4}a).
We vary the cone apex angle ($\alpha$) and dipole ($\lambda_0 = 650$ nm) location in the vertical direction while keeping the PMMA thickness ($h= 620$ nm) and cone height ($h_c = 170$ nm ) fixed as in the optimized structure discussed later. (figure \ref{fig:4}b-d)
It can be seen from figure \ref{fig:4}b that locating the dipole within a few nanometers from the cone tip is essential in achieving high enhancement factors.
The dipole height is fixed at $4$ nm due to fabrication limitations in subsequent calculations.
Figure \ref{fig:4}c displays the redshift of the longitudinal plasmonic resonance as the apex angle is reduced as discussed above. 
An important aspect is that the enhancements are broadband with FWHM $\sim80$ nm which is essential for broadband room-temperature emitters and fits well with the the broadband emission redirection offered by the bullseye nanoantenna. 
Furthermore we find that the antenna quantum efficiency is best for larger nanocones as has been reported previously (figure \ref{fig:4}d). \cite{DAgostino2013DipoleNanocones,Mohammadi2010FluorescenceAntenna}
\begin{figure}[t!]
\centering{\includegraphics[width=0.5 \textwidth]{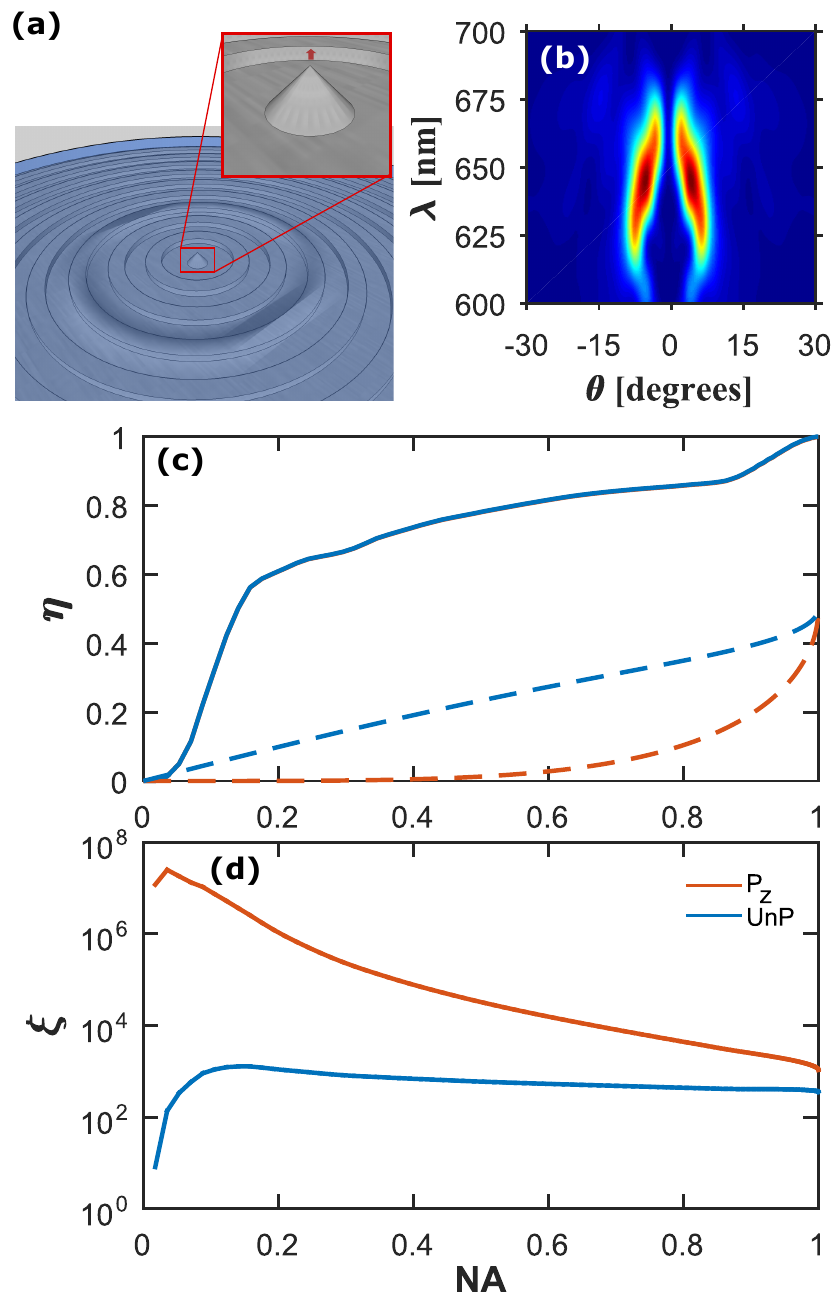}}
\caption{(a) Schematic for the a silver nanocone embedded in a bullseye nanoantenna. (b) Angular intensity spectrum $I(\theta,\lambda)$  for a dipole emitter ($\lambda_0=650$ nm) in a composite bullseye nanocone antenna. Collection efficiency   (c) and brightness enhancement (d) of a dipole emitter in the optimized composite structure for the vertical dipole ($P_z$) and unpolarized dipole ($UnP$). The dashed lines represent the values for dipoles in free space }
\label{fig:5}
\end{figure}

Next we place the nanocone at the center of the bullseye nanoantenna. 
A close-up view of the structure is shown schematically in figure \ref{fig:5}a with a zoom-in display of the nanocone in the inset.
The different dipole orientation and effective dipole height force a new optimization for both collection efficiency and brightness enhancement factor. 
We thus use the same figure of merit as in section 3 which is the far-field brightness for an NA of 0.22 (i.e. $\Phi_z(0.22)$). 
The optimization is carried out for a dipole emitter at central wavelength $\lambda_0 = 650$ nm and which is oriented along the cone axis ($P_z$).
The optimization clearly pushes towards thicker polymer layers and lower groove heights due to the preferential coupling of the nanocone-emitter sub-system into the $TM$ modes of the waveguide layer as discussed in section 3.
The parameters for the bullseye nanoantenna are: normalized central radius $R/\lambda_0 = 0.99 $; normalized grating period $\Lambda/\lambda_0 = 0.86$; normalized polymer thickness $h/\lambda_0 = 0.95$; and normalized groove heights and widths equal to $0.167$ and $0.500$ respectively.
A nanocone of height $h_c = 170$ nm and apex angle $\alpha=50^o$ is placed at the center of the bullseye structure and the dipole emitter is located 4 nm above the tip of the nanocone. 
The far-field angular intensity distribution is displayed in figure \ref{fig:5}b featuring a broad resonance and narrow angular emission pattern. 
At resonance ($\lambda=650$ nm) the angular pattern has a prominent angular ring at an angle of $\sim 4.5^o$.
As shown in figure \ref{fig:5}c this pattern results in a collection efficiency into an $NA=0.22$ ($0.5$)  of around $61 \%$ ($78 \%$) respectively for both the $P_z$ and unpolarized dipole orientations. 
Due to high enhancement of one dipole orientation over the other the flux of an unpolarized emitter acquires the dependency of the $P_z$ dipole with an overall flux of about $1/3$ that of the $P_z$ dipole.  
Thus the collection efficiency of the unpolarized and $P_z$ dipoles are essentially identical. 

A brightness enhancement factor into these NA's are $\sim1000$ ($\sim600$) for an unpolarized emitter (figure \ref{fig:5}d). The extremely high brightness enhancement at low NA for the $P_z$ dipole is due to the extremely poor collection efficiency of the corresponding free dipole (see figure \ref{fig:5}c).  
Moreover the overall radiative enhancement factor is $\sim 240$ with an antenna efficiency of around $31\%$.

The implementation of this design will vary depending on the particular emitter in question. 
A particularly interesting application would be for NQDs since the ability to bond these nanocrystals to the tips of the nanocone has been developed. \cite{Fulmes2015Self-alignedNanostructures,Meixner2015} In particular, the biexciton quantum efficiency will be increased due to the low intrinsic quantum efficiency of the biexciton emission (see figure \ref{fig:1}a and equation \ref{eq:QY}),\cite{Matsuzaki2017StrongAntenna}, giving rise to a potential bi-photon source.

For sources where the single-photon purity is retained even when the decay rate is enhanced such as color centers in diamond, this design will permit the fabrication of a bright and efficient single photon source.
Such a source will improve  quantum key distribution systems which currently use faint laser pulses with an average photon number of $\left<n\right>\approx0.5$ per pulse. 
An ideal single photon source will have $\left<n\right>\approx1$ which requires both a high single photon purity and high collection efficiency ( $>0.5$) \cite{Aharonovich2016Solid-stateEmitters}. 
Furthermore the emission rate must be in the GHz regime to compete with the current technology which requires high radiative enhancement factors. 
Since SiV centers already have intrinsic lifetimes on the order of $\sim 1$ ns and display good single photon purities \cite{Li2016NonblinkingNanodiamonds} it is not long shot to satisfy these requirements using the composite bullseye-nanocone antenna.
Small nanodiamonds containing SiV centers may be placed in the proximity of the nanocone by a scanning probe tip using a pick and place technique \cite{Schell2011ADevices}. 

\section{Conclusion}
Our new design and FDTD calculations suggest an optimized bullseye nanoantenna with over $82 \%$ collection efficiency into an NA of a multimode fiber and a corresponding source brightness that is more than 10 fold that of a dipole of free space.
 We have also investigated a composite structure that combines both high collection efficiency ($\eta(NA=0.22)=0.61$ ) and high radiative enhancement ($\Gamma_z^r=240$) using a silver nanocone at the center of a bullseye nanoantenna. 
 
These results are especially promising for the production of a high-rate, room-temperature on-chip single photon sources that can be directly coupled to optical fibers without any additional optics.
This structure can be also used for example along with color centers in diamond (e.g. SiV and NV centers) enabling applications spanning various fields including quantum communication and magnometric sensing \cite{Schroder2016QuantumInvited}.
It is important to emphasize that these structures are not limited to any specific quantum emitter and can be applied universally, limited only by challenges in fabrication. 

\section*{Acknowledgements}
This work is supported in parts by The Einstein Foundation Berlin; The U.S. Department of Energy: Office of Basic Energy Sciences,
Division of Materials Sciences and Engineering; The European Cooperation in Science
and Technology through COST Action MP1302 Nanospectroscopy; The Ministry of Science
and Technology, Israel.
\bibliography{Mendeley.bib} 
\bibliographystyle{unsrt}

\pagebreak

\subfile{supplementary.tex}

\end{document}

%% file: supplementary.tex
\textbf{\LARGE 
\begin{center}
Supplementary Information
\end{center}}

\noindent\textbf{\Large Quantum emitters coupled to circular nanoantennas for high brightness quantum light sources}

Hamza A. Abudayyeh $^1$ and Ronen Rapaport $^{1,2}$

$^1$ Racah Institute for Physics, and
$^2$ the Applied Physics Department, The Hebrew University of Jerusalem, Jerusalem 9190401, Israel
\nopagebreak
\\
\\
\\

\setcounter{section}{0}
\setcounter{figure}{0}
\setcounter{equation}{0}

\section{Modes of metal dielectric air (MDA) waveguide}

\begin{figure}[t!]
\centering{\includegraphics[width=0.5\textwidth]{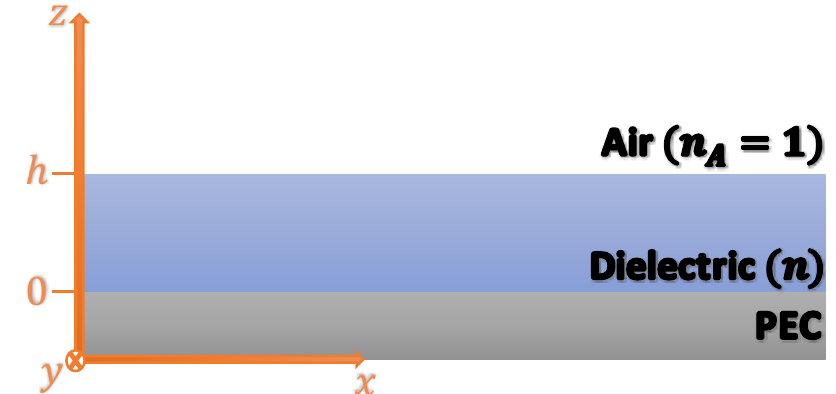}}
\caption{Diagram of metal dielectric air waveguide.}
\label{MDAfig}
\end{figure}

The system considered here is shown in figure \ref{MDAfig}. 
This structure is a special form of the three layer waveguide structure with the substrate as a perfect electric conductor and the cladding is air. 
The solutions of the wave equations that we are interesting in are those that result in a propagating waveguide mode (i.e. in the x direction). 
It is also assumed that the waveguide is homogeneous in the y-direction. 
Therefore the spatial dependence of the electromagnetic fields will have the following form:

\begin{equation}
\label{eq:VecE}
\begin{aligned}
\vec{E}(x,y,z) &=\vec{e}(z) \exp(i\beta x) 
\\
\vec{H}(x,y,z) &=\vec{h}(z) \exp(i\beta x) 
\end{aligned}
\end{equation}
Substituting equations\ref{eq:VecE} into Maxwell's equations results in two sets of solutions for the transverse electric (TE) and transverse magnetic (TM) polarizations.

\subsection*{TE polarization}
For this polarization the nonzero field components are $e_y$, $h_x$ and $h_z$. 
Since we are looking for propagating modes in the waveguide we assume that the fields are evanescent in the air layer. 
Furthermore the parallel electric field must vanish at the PEC surface. 
Thus it is clear that the solutions must take the form of:
\begin{equation}
e_y(z)= 
\begin{cases}
A\sin(\kappa z),  &  0\leq z \leq h \\ 
B\exp\left(-\gamma (z-h)\right),  &  z \geq h
\end{cases}
\end{equation}
Assuming harmonic time dependence, $\kappa $ and $\gamma$ must satisfy :
\begin{equation}
\label{eq:kconservation}
\begin{aligned}
\beta^2  +\kappa^2  &=\frac{n^2\omega^2}{c^2} \\
\beta^2  -\gamma^2 &= \frac{\omega^2}{c^2}
\end{aligned}
\end{equation}
The parallel electric and magnetic fields need to be continuous at the dielectric air boundary which results in the following equations:
\begin{equation}
\label{eq:BCTE}
\begin{aligned}
A\sin(\kappa h)&=B
\\ A\kappa \cos(\kappa h)& = -\gamma B
\end{aligned}
\end{equation}
By dividing these equations we reach to the following transcendental equation:
\begin{equation}
\label{eq:modeTE}
\tan(\kappa h)=-\frac{\kappa}{\gamma}
\end{equation}
By solving equations  \ref{eq:kconservation} and \ref{eq:modeTE} simultaneously one obtains the dispersion relation ($\omega (\beta)$). Clearly this would lead to an infinite number of modes each having a cutoff that occurs when the field is no longer confined to the waveguide layer (i.e. $\gamma \rightarrow 0 $). Therefore the cutoff condition occurs when $\tan(\kappa h)= -\infty$. This can be written in terms of the free-space wavelength $\lambda_0= 2\pi c/\omega$  and the waveguide thickness as: 
\begin{equation}
\label{eq:cutTE}
\frac{h_{cutoff}^{(\nu)}}{\lambda_0}= \frac{2\nu+1}{4\sqrt{n^2 -1}} 
\end{equation}
where $\nu = 0,1,2,... $ is the mode number and $h_{cutoff}^{(\nu)}$ is the minimum waveguide thickness that would support the TE$_{\nu}$ mode.
The electric field of the TE$_\nu$ mode can therefore  be given as:
\begin{equation}
\label{eq:ETe}
\resizebox{0.4\textwidth}{!}{$
E_y^{(\nu)}(x,z)= 
\begin{cases}
E_0\sin(\kappa_\nu z)\exp(i \beta_\nu x),  &  0\leq z \leq h \\ 
E_0 \sin(\kappa_\nu d)\exp\left(-\gamma_\nu (z-h)\right)\exp(i \beta_\nu x),  &  z \geq h
\end{cases}$}
\end{equation}
, and the magnetic field may be found by applying Maxwell's equations.
\subsection*{TM polarization}
For this polarization the non-vanishing fields are $h_y$, $e_x$, and $e_z$ . Therefore it is easier to solve the equations in terms of the magnetic field $h_y$ and then apply Maxwell's curl equations to find the electric field components.  As before we assume a confined propagating field which leads to the ansatz: 
\begin{equation}
h_y(z)= 
\begin{cases}
A\cos(\kappa z),  &  0\leq z \leq h \\ 
B\exp\left(-\gamma (z-h)\right),  &  z \geq h
\end{cases}
\end{equation}
where the cosine was chosen to force the tangential electric field ($e_x$) to vanish at the conductor surface. The parameters $\kappa$ and $\gamma$ are still related to the propagation constant $\beta$ by equation \ref{eq:kconservation}. By applying the continuity of the tangential electric and magnetic fields at the dielectric air interface we reach to the following conditions: 
\begin{equation}
\label{eq:BCTM}
\begin{aligned}
A\cos(\kappa h)&=B
\\ \frac{A}{n^2}\kappa \sin(\kappa h)& = \gamma B
\end{aligned}
\end{equation}
which yields the transcendental equation: 
\begin{equation}
\label{eq:modeTM}
\tan(\kappa h)=\frac{n^2\gamma}{\kappa}
\end{equation}
Again a mode is cutoff when $\gamma \rightarrow 0 $, or $\tan (\kappa h)=0$. This yields the following condition:
\begin{equation}
\label{eq:cutTM}
\frac{h_{cutoff}^{(\nu)}}{\lambda_0}= \frac{\nu}{2\sqrt{n^2 -1}} 
\end{equation}
with $\nu = 0,1,2,...$ corresponds to the TM$_\nu$ mode. One important difference is that the TM$_0$ mode has no cutoff and therefore is always present regardless of the thickness or wavelength considered.  

We can get the electric field components by applying the curl equations to obtain: 
\begin{equation}
\label{eq:ETM}
\resizebox{0.4\textwidth}{!}{$
\begin{aligned}
E^{(\nu)}_x(x,z)=
\begin{cases}
E_0 \sin(\kappa_\nu z)\exp(i \beta_\nu x),  &  0\leq z \leq h \\ 
E_0 \sin(\kappa_\nu h)\exp(-\gamma_\nu (z-h))\exp(i \beta_\nu x),  &  z \geq h
\end{cases}
\\
E^{(\nu)}_z(x,z)=
\begin{cases}
\frac{-i E_0 \beta_\nu}{\kappa_\nu} \cos(\kappa_\nu z)\exp(i \beta_\nu x),  &  0\leq z < h \\ 
\frac{i E_0 \beta_\nu}{n^2} \sin(\kappa_\nu h)\exp(-\gamma_\nu (z-h))\exp(i \beta_\nu x),  &  z > h
\end{cases}
\end{aligned}
$}
\end{equation}
\section{Incoherent unpolarized dipole sources}
Here we will discuss how the fields and flux of an unpolarized dipole source may be calculated. 
The field by an unpolarized source may be calculated by an incoherent superposition of the fields of randomly oriented oscillating dipoles, i.e.:
\begin{equation}
\label{eq: normincoherent}
\left<|\Psi|^2\right> = \frac{1}{4\pi} \int\int |\Psi(\theta',\phi')|^2 \sin(\theta')d\theta' d\phi' 
\end{equation}
where $\Psi$ is a general vector field that can represent the electric or magnetic field and $<\cdots >$ signifies an average over all dipole oreientations . $\theta'$ and $\phi'$ are the polar and azimuthal angles of the dipole and are not to be confused with the polar and azimuthal angles of the position vector $\vec{r}$ ($\theta$,$\phi$). It is worth noting that all the  fields in this section have implicit dependence on $\vec{r}$ which we dropped from the notation.  
\begin{figure}[t!]
\centering{\includegraphics[width=0.5 \textwidth]{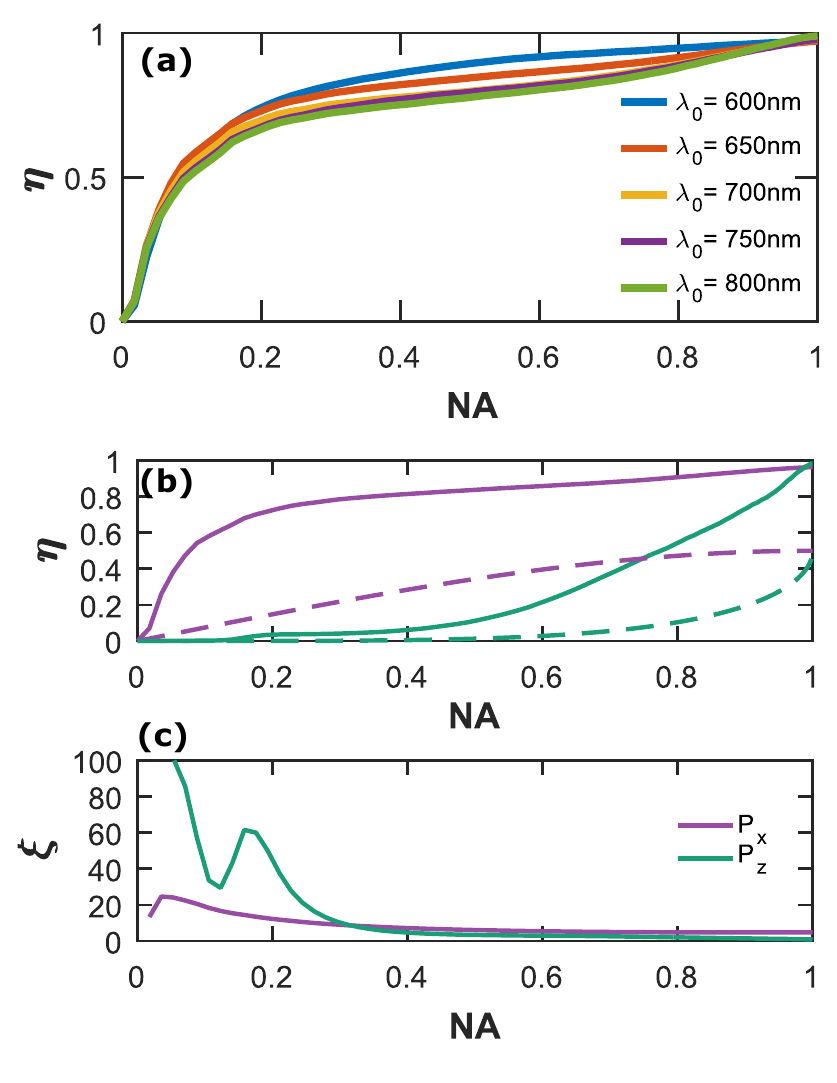}}
\caption{(a) Collection efficiency for the structure scaled to different central wavelengths. Collection efficiency ($\eta$) (b) and Brightness enhancement ($\xi$)  (c) for the optimized structure for dipoles oriented along the x and z axes (solid curves) . The dashed lines represent the curves for dipoles in free space.} 
\label{fig:Supp2}
\end{figure}
Such a randomly oriented dipole can in general be decomposed into three orthogonal dipoles $\vec{P}_x$, $\vec{P}_y$ , and $\vec{P}_z$ oriented along the axes of some coordinate system as follows: 
\begin{equation}
\vec{P}= \vec{P}_x \sin(\theta')\cos(\phi') + \vec{P}_y \sin(\theta')\sin(\phi')+ \vec{P}_z \cos(\theta') 
\end{equation}
We can therefore write the  field of an unpolarized source as:
\begin{equation}
\label{eq: randomdipole}
\vec{\Psi}(\theta',\phi')= \vec{\Psi}_x\sin(\theta')\cos(\phi')+ \vec{\Psi}_y\sin(\theta')\sin(\phi')+ \vec{\Psi}_z\cos(\theta')
\end{equation}
where $\vec{\Psi}_i$ represents the  field due to the dipole $\vec{P_i}$ .  Substituting equation \ref{eq: randomdipole} into equation \ref{eq: normincoherent} and using some integral identities gives the following relation: 
\begin{equation}
\label{eq: averagedfield}
\left<|\Psi|^2\right> = \frac{1}{3}\left( |\Psi_x|^2+|\Psi_y|^2+|\Psi_z|^2 \right)
\end{equation}
This derivation assumes that the orientation of the dipole is completely random i.e. $P_i$ and thus $\Psi_i$ have no dependence on $\theta'$ or $\phi'$ .  Equation \ref{eq: averagedfield} is equally valid for the electric and magnetic field and is therefore valid for the flux: 
\begin{equation}
\label{eq: averagedflux}
\left<F\right> = \frac{1}{3}\left( F_x+F_y+F_z \right)
\end{equation}
For a  system with azimuthal symmetry this reduces to:
\begin{equation}
\label{eq: averagedfluxazimuthal}
\left<F\right> = \frac{2}{3} F_x+\frac{1}{3}F_z 
\end{equation}\section{Circular periodic ('Bullseye') nanoantenna with a central cavity}
In this section we show supplementary results for section 3 in the main text. 

\begin{figure}[t!]
\centering{\includegraphics[width=0.5 \textwidth]{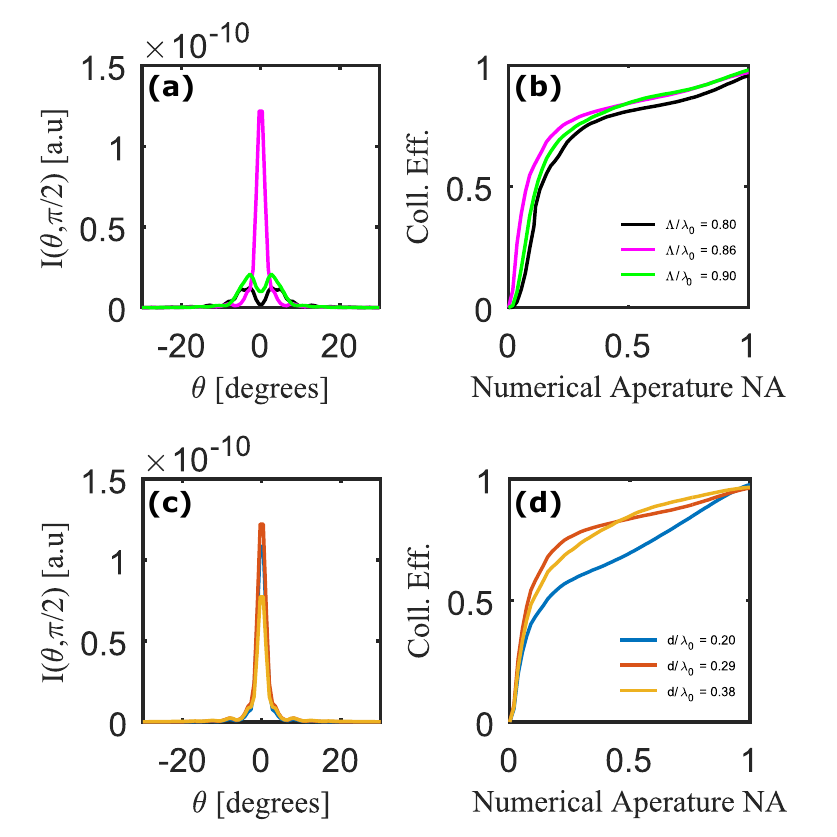}}
\caption{Intensity angular distribution in the yz-plane (a and c) and the collection efficiency (b and d) for various values of the period (a and b) and the dipole height (c and d).  }
\label{fig:Supp3}
\end{figure}

\subsection*{Device geometrical parameters scalability with emitter wavelength:}
One important feature that the design of the nanoantenna possesses is scalability of its geometrical parameters with the emitter's wavelength. 
For this reason we chose in the main text to normalize lengths to the emitter's central wavelength to make the reported parameters general for all relevant emission wavelengths.
In order to display this we plot in figure \ref{fig:Supp2}a the collection efficiency for various central wavelengths between $600$ and $800$ nm where we scale the nanoantenna dimensions to the central wavelength. 
The figure clearly shows that the differences between the various central wavelengths is small and that the overall trend is the same for all wavelengths. 
The differences can be mainly attributed to the wavelength dispersion of the metal optical properties. 
The dispersion of the dielectric in the wavelength range considered can be neglected. 
\subsection*{Device emission dependence on dipole orientation}
We complement the discussion about the dipole orientation in the main text here by displaying the collection efficiency and brightness enhancement (defined in equations 6 and 7 in the main text) for the $P_z$ dipole in figure \ref{fig:Supp2}b and c respectively. 
As compared to the $P_x$ orientation the $P_z$ dipole is both suppressed ($F_z^r \approx 0.45 $ ) and the collection efficiency is low for low numerical apertures. 
This emphasizes the point made in the main text that an unpolarized emitter will emit preferentially with a dipole oriented along the x-axis especially at low NA. 
\subsection*{Device emission dependence on geometrical parameters}
To continue the parameter dependence analysis in the main text, here we discuss the effect of changing the periodicity of the circular grating and the location of the dipole emitter in the polymer layer.
Figure \ref{fig:Supp3} represent the intensity angular distributions and collection efficiency for different values of these parameters. 
The parameters are detuned above an below the optimized parameters ($\Lambda/\lambda_0 =0.86$ and $d/\lambda_0 = 0.29$ ). 
The effect of detuning the periodicity leads to a mismatch between the waveguide propagation constant and the grating Bragg wavevector causing  the formation of sidebands. 
On the other hand changing the dipole location will lead to lower coupling between the dipole emission and the waveguide mode. 
This will have the primary effect of lowering the radiative enhancement as can be seen from figure \ref{fig:Supp3}c.
It will slightly change the collection efficiency of the antenna (see figure \ref{fig:Supp3}d).